\documentclass[epj]{webofc}
\usepackage[varg]{txfonts}   % Web of Conferences font
\usepackage{graphicx}

\usepackage{hyperref}
\usepackage{float}
\usepackage{epsfig}
\usepackage{graphicx}
\usepackage{mathptmx}      % use Times fonts if available on your TeX system
\usepackage{latexsym}
\usepackage{longtable}
\usepackage{dcolumn}
\usepackage{natbib}
\usepackage{comment}

\wocname{EPJ Web of Conferences}
\woctitle{CONF12}
\begin{document}
\selectlanguage{english} 
\title{The falsification of Chiral Nuclear  Forces}
%
% subtitle (optional, strongly discouraged)

\author{E. Ruiz
  Arriola\inst{1}\fnsep\thanks{\email{earriola@ugr.es}}~\footnote{Speaker
at XIIth Conference on Quark Confinement and the Hadron Spectrum. \\ Work
      supported by Spanish Ministerio de Economia y Competitividad and
      European FEDER funds (grant FIS2014-59386-P), the Agencia de
      Innovacion y Desarrollo de Andalucia (grant No. FQM225), the
      U.S. Department of Energy by Lawrence Livermore National
      Laboratory under Contract No. DE-AC52-07NA27344, U.S. Department
      of Energy, Office of Science, Office of Nuclear Physics under
      Award No. DE-SC0008511 (NUCLEI SciDAC Collaboration)} \and 
  J. E. Amaro\inst{1}
%\fnsep\thanks{\email{amaro@ugr.es}} 
\and 
R. Navarro  Perez\inst{2}
%\fnsep\thanks{\email{navarroperez1@llnl.gov}}
}
\institute{Departamento de F\'{\i}sica At\'omica, Molecular y Nuclear
  and \\ Instituto Carlos I de F{\'\i}sica Te\'orica y Computacional,
  Universidad de Granada \\ E-18071 Granada, Spain \and
Nuclear and Chemical Science Division, Lawrence Livermore
  National Laboratory\\ Livermore, California 94551, USA
          }

\abstract{Predictive power in theoretical nuclear physics has been a
  major concern in the study of nuclear structure and reactions. The
  Effective Field Theory (EFT) based on chiral expansions provides a
  model independent hierarchy for many body forces at long distances
  but their predictive power may be undermined by the regularization
  scheme dependence induced by the counterterms and encoding the short
  distances dynamics which seem to dominate the uncertainties. We
  analyze several examples including zero energy NN scattering or
  perturbative counterterm-free peripheral scattering where one would
  expect these methods to work best and unveil relevant systematic
  discrepancies when a fair comparison to the Granada-2013 NN-database
  and partial wave analysis (PWA) is undertaken.}
\maketitle
\section{Introduction}
\label{intro}

Nuclear Physics has always been characterized by the fact that
experiment is much more precise than theory. For nuclear masses one
has $ \Delta M (Z,N)^{\rm exp} < 1 {\rm KeV} \ll \Delta M (Z,N)^{\rm
  th} $ but it is unclear what the theoretical uncertainty
is. Traditionally, the theoretical and reductionist predictive power
flow is expected to be from light to heavy nuclei form a Hamiltonian 
with multinucleon forces 
\begin{equation}
H(A) = T + V_{2N} + V_{3N} + V_{4N}+ \dots \quad \to \quad H(A)\Psi_n = E_n(A) \Psi_n \, .  
\end{equation}
In the absence of {\it ab initio} determinations, phenomenological
$V_{nN}$ interactions are adjusted to NN scattering and light nuclei
binding energies. The chiral approach, originally suggested by
Weinberg in 1990~\cite{Weinberg:1990rz} (see
e.g. \cite{Bedaque:2002mn,Epelbaum:2008ga,Machleidt:2011zz} for
reviews) to nuclear forces provides a power counting in terms of the
pion weak decay constant $f_\pi$, with the appealing feature of
systematically providing a hierarchy
\begin{equation}
V_{2N}^{\chi} \gg V_{3N}^{\chi} \gg V_{4N}^{\chi} \gg \dots 
\end{equation}
Because the pion mass is so small, chiral approaches are unambiguous
at long distances via $1\pi$,$2\pi$,$3\pi,\dots$ exchanges for
relative distances above a short distance cut-off $r_c$, $V^{n \pi}
(r_c) \sim e^{- n r_c m_\pi }$. For instance, NN chiral potentials,
constructed in perturbation theory, are universal and contain chiral
constants $c_1,c_3,c_4, \dots$ which can be related to $\pi N$
scattering~\cite{Bedaque:2002mn,Epelbaum:2008ga,Machleidt:2011zz}. At
long distances we have
\begin{equation}
V^{\chi}_{NN}(r) =  V^{\pi}_{NN}(r) + V^{2 \pi}_{NN}(r) + V^{3 \pi}_{NN}(r) + \dots  \qquad r \gg r_c \, , 
\end{equation}   
whereas they become singular at short distances 
\begin{equation}
V^{\chi}_{NN}(r) =  \frac{a_1}{f_\pi^2 r^3} +  \frac{a_2}{f_\pi^4 r^5} + 
 \frac{a_3}{f_\pi^6 r^7} +  \dots  \qquad r \ll r_c \, , 
\end{equation} 
and some regularization must be introduced in {\it any} practical
calculation.  Thus, they trade the ``old'' model dependence for the
``new'' regulator dependence.  What is the best theoretical accuracy
we can get within ``reasonable'' cut-offs ? What is a reasonable
cut-off ? Can the short distance piece be organized as a power
counting compatible with the chiral expansion of the long distance
piece ?

The huge effort which has been carried out over the last 25 years
ellapsed since the seminal work of Weinberg, harvesting over 1000
citations, proves the computational feasibility of the chiral nuclear
agenda requiring large scale calculations and many CPU computing
hours. Here, we depart from the main streamline and wonder if chiral
nuclear forces can be falsified or validated and, if yes, if they are
useful for nuclear structure applications from the point of view of
the predictive power.

Of course, all this has to do with proper assessment and evaluation of
uncertainties of any sort and in particular in the NN interaction. Our
original and simple estimates~\cite{NavarroPerez:2012vr,Perez:2012kt}
of $\Delta B^{\rm th}/A \sim 0.5 \, {\rm MeV}$ has been upgraded in
Ref.~\cite{Perez:2014waa} to be enlarged to $\sim 2 \, {\rm
  MeV}$. These crude estimates are not far from the recent uncertainty
analysis and order-by-order optimization of chiral nuclear
interactions~\cite{Carlsson:2015vda} including three-body forces where
it is found $\Delta B^{\rm th} ( ^{16}{\rm O}) /16 \sim 4 {\rm MeV}$.
Most of the uncertainty stems from the cut-off variation within a
``reasonable'' range, and is much worse than the ancient
Weiszacker semiempirical mass formula, where $\Delta B_{\rm sem}/A
\sim 0.1 {\rm MeV}$.  If confirmed, it would be a rather pesimistic
scenario for the chiral approach to nuclear structure pioneered by
Weinberg. Motivated by this alarming possibility we have paid
dedicated attention in the last five years to the issue of NN
uncertainties~\cite{Perez:2014kpa,Perez:2014waa}. Here we focus on
$V_{NN}^{2\pi}$, corresponding to chiral $2\pi$ exchange ($\chi$TPE).

\section{Validation and Falsification: Frequentist vs Bayesian}

From our point of view, making first a fair statistical treatment is a
prerequisite to credibly aim at any precision goal in low energy
nuclear physics where information is extracted by fits. We remind the
fact that least squares $\chi^2$-fitting any (good or bad) model to
some set of data is always possible and corresponds to just minimizing
a distance between the predictions of the theory and the experimental
measurements.  How can we disentangle between {\it true} and {\it
  false} models?.

The well-known statistical approach, to which we stick, provides one
probabilistic answer and depends on the number of data, $N_{\rm Dat}$,
the number of parameters determined from this data, $N_{\rm Par}$, and
the nature of experimental uncertainties.  The natural question is:
What is the probability that given the data the theory is correct ?
This corresponds to the Bayesian approach which requires some {\it a
  priori} expectations on the goodness of the theory regardless of the
data and is dealt with often by augmenting the experimental $\chi_{\rm
  exp}^2$ with an additive theoretical contribution $\chi_{\rm th}^2$.
However, it can be proven that when $N_{\rm Dat} \gg N_{\rm Par}$ one
can ignore these a priori expectations since $\chi^2_{\rm exp} \sim
N_{\rm Dat} \gg \chi_{\rm th}^2 \sim N_{\rm Par}$ and proceed with the
frequentist approach where just the opposite question is posed: what
is the probability of data given the model ?.~\footnote{One could stay
  Bayesian if some relative weighting of $\chi_{\rm exp}^2$ and
  $\chi_{\rm th}^2$ is implemented
  (see~\cite{Ledwig:2014cla,Perez:2016vzj} and references therein).}
In our analysis below, where we have $N_{\rm Dat} \sim 8000$ and
$N_{\rm Par} \sim 40$, we expect no fundamental differences. We thus
simply ask: what is the probability $q$ that the {\it the model is
  false} ?. The p-value is $p=1-q$ and if $p$ is smaller than a
predetermined confidence level we will not trust the model and we will
declare it to be false. Note that 1) we can never be sure that the
model is true and 2) any experiment can be right if errors are
sufficiently large and the theory cannot be falsified. This said,
$p=0.68$ when $\chi^2/\nu = 1 \pm \sqrt{2/\nu}$ with $\nu=N_{\rm
  Dat}-N_{\rm Par}$.

In general we expect discrepancies between theory and data and,
ideally, if our theory is an approximation to the true theory we
expect the optimal accuracy of the truncation to be comparable with
the given experimental accuracy and both to be compatible within their
corresponding uncertainties (see \cite{Wesolowski:2015fqa} for a
Bayesian viewpoint).  If this is or is not the case we validate or
falsify the approximated theory against experiment and declare theory
and experiment to be compatible or incompatible respectively. Optimal
accuracy, while desirable, is not really needed to validate the
theory. In the end largest errors dominate regardless of their
origin; the approximated theory may be valid but inaccurate.

How should the discrepancies or residuals be interpreted ? Statistics
has the obvious advantage that if we have no good reasons to suspect
the theory we can test if residuals behave as, often gaussian, fluctuations and
determine a confidence interval for fitting parameters within these
fluctuations.

\section{Fitting and selecting data form coarse grained potentials}

The NN scattering amplitude has 5 independent complex components for
any given energy, which must and can be determined from a complete set
of measurements involving differential cross sections and polarization
observables. From this point of view it is worth reminding that phase
shifts obtained in PWA are {\it not} data by themselves unless a
complete set of 10 fixed energy and angle dependent measurements have
been carried out, a rare case among the bunch of existing 8000 np+pp
scattering data below $350 \, {\rm MeV}$ LAB energy and which
corresponds to a maximal CM momentum of $p_{\rm CM}^{\rm max}= 2 {\rm
  fm}^{-1}$. In order to intertwine all available, often incomplete
and partially self-contradictory, information some energy dependence
interpolation is needed.  We assume a potential approach inspired by a
Wilsonian point of view where we take a grid of equidistant radial
``thick'' points in coordinate space separated by the finite
resolution given by the shortest de Broglie wavelength, $\Delta r =
\hbar /p_{\rm CM}^{\rm max} \sim 0.6 {\rm fm}$ up to the radius $r_c=
3 \, {\rm fm}$, above which charge dependent $1\pi$ exchange gives the entire
strong contribution. The counting of parameters~\cite{Perez:2013cza}
yields about 40 ``thick'' points, which can be represented by
delta-shells (DS)~\cite{NavarroPerez:2012qf} as originally proposed by
Avil\'es~\cite{Aviles:1973ee}. The whole procedure needs long distance
electromagnetic and relativistic contributions such as Coulomb, vacuum
polarization and magnetic moments interactions.  This approach allows
to select the largest self-consistent existing NN database with a
total of 6713 NN scattering data driven by the coarse grained
potential~\cite{Perez:2013jpa,Perez:2014yla} with the rewarding
consequence that statistical uncertainties can confidently be
propagated.~\footnote{This resulting Granada-2013 database
  (\url{http://www.ugr.es/~amaro/nndatabase/}) will be used in our
  discussion.} Precise determinations of chiral coefficients,
$c_1,c_3,c_4$~\cite{Perez:2014bua,Perez:2013za}, the isospin breaking
pion-nucleon~\cite{Perez:2016aol,Arriola:2016hfi}, and the
pion-nucleon-delta~\cite{Perez:2014waa} coupling constants have been
made.

\begin{table}
  \caption{\label{tab:Cut-offdependence} 
Fits of chiral TPE potentials depending on the cutoff radius and the maximum
fitting energy~\cite{Perez:2013za}.}
%Different delta-shell
%    potentials with chiral two pion exchange tail fitted to the
%    self-consistent Granada database of~\cite{Perez:2013jpa}. The
%    first line corresponds to the potential presented
%    in~\cite{Perez:2013oba} and the second line to the potential of
%    this work. The chiral constants of the fourth line were taken from
%    a Ref.~\cite{Ekstrom:2013kea, Ekstrom:2014dxa} and used as fixed
%    values during the $\chi^2$ minimization with respect of the
%    delta-shell parameters. Highest counterterm column indicates the
%    maximum angular momentum where at least one delta-shell strength
%    coefficient is non-vanishing.}
%  \begin{ruledtabular}
    \begin{tabular*}{\columnwidth}{@{\extracolsep{\fill}} l c D{.}{.}{3.6} 
        D{.}{.}{2.5} D{.}{.}{1.4}  c c}
      Max $T_{\rm LAB}$ & $r_c$ & 
      \multicolumn{1}{c}{$c_1$} & 
      \multicolumn{1}{c}{$c_3$} & 
      \multicolumn{1}{c}{$c_4$} & Highest  & $\chi^2/\nu$ \\
      MeV & fm & 
      \multicolumn{1}{c}{GeV$^{-1}$} & 
      \multicolumn{1}{c}{GeV$^{-1}$} & 
      \multicolumn{1}{c}{GeV$^{-1}$} & counterterm & \\
      \hline \noalign{\smallskip}
      350 & 1.8 & -0.4(11) & -4.7(6)  & 4.3(2) & $F$ & 1.08 \\
      350 & 1.2 & -9.8(2)  &  0.3(1)  & 2.84(5)& $F$ & 1.26 \\
      125 & 1.8 & -0.3(29) & -5.8(16) & 4.2(7) & $D$ & 1.03 \\
      125 & 1.2 & -0.92    & -3.89    & 4.31   & $P$ & 1.70 \\
      125 & 1.2 & -14.9(6) & 2.7(2)   & 3.51(9)& $P$ & 1.05 \\
    \end{tabular*}
%  \end{ruledtabular}
\end{table}

\section{Chiral Fits, peripheral waves}

The questions on the cut-of $r_c$ raised above were answered 
by separating the potential as follows~\cite{Perez:2013za}
\begin{equation}
V(r) = V_{\rm short} (r) \theta (r_c - r ) + V_{\rm long}^{\chi} (r)
\theta (r- r_c) \, , \qquad V_{\rm short}= \frac1{2\mu}\sum_{n} \lambda_n \delta (r-r_n) \, , 
\end{equation}
with $r_n = n \Delta r$. Several fits varying $r_c$ and $E_{\rm
  LAB}^{\rm max}$ were performed. The results were checked to be
statistically consistent and are summarized in
Table~\ref{tab:Cut-offdependence}. It is striking that $D$-waves,
nominaly N3LO and forbidden by Weinberg chiral counting at N2LO, are
indispensable !. Furthermore, data and N2LO do not support $r_c <
1.8{\rm fm}$, while several
$\chi$-potentials~\cite{Gezerlis:2014zia,Piarulli:2014bda} take
$r_c=0.9-1.1 {\rm fm}$ as ``reasonable''.

An alternative way of checking the failure of the power counting is
provided by a deconstruction argument~\cite{Perez:2013za}. This
corresponds to determine under what conditions are the short distance
phases $\delta_{\rm short} $, .i.e. the phase shifts stemming solely
from $V_{\rm short}$ compatible with zero within uncertainties,
i.e. $|V_{\rm short}| < \Delta V$ ?.  This corresponds to check what
partial waves fullfill $ | \delta_{\rm short} | \le \Delta \delta_{\rm
  stat}$ when $r_c = 1.8 \, {\rm fm} $. Unfortunately, this does not
work for D-waves, supporting the previous conclusions.

The long distance character of $\chi$TPE makes peripheral phases
(large angular momentum) to be suitable for a perturbative comparison
{\it without}
counterterms~\cite{Kaiser:1997mw,Kaiser:1998wa,Entem:2014msa}. However,
one should take into account that 1) peripheral phases can only be
obtained from a complete phase shift analyses and 2) their
uncertainties are tiny~\cite{Perez:2013jpa}. The analysis of
\cite{Entem:2014msa} just makes an eyeball comparison which looks
reasonable but the agreement was not quantified.~\footnote{This was
  done using the SAID database (\url{http://gwdac.phys.gwu.edu/}), a
  $25\sigma$ incompatible fit with $p \ll 1$(see e.g.~\cite{Perez:2014waa}).}
We find that peripheral waves predicted by 5th-order chiral
perturbation theory are not consistent with the Granada-2013
self-consistent NN database 

\begin{equation}
|\delta^{\rm Ch,N4LO}- \delta^{\rm PWA}| >  \Delta \delta^{\rm PWA, stat}  \, . 
\end{equation}
Sometimes we get even $3\sigma$ discrepancies.  
More details on this
peripheral analysis will be presented elsewhere. Of course, one may
thing that 125 MeV is too large an energy. We find that when we go
down to 40 MeV, the $\chi$TPE potential becomes invisible being compatible
with zero~\cite{Amaro:2013zka,Perez:2013cza}. 

The chiral potential (including $\Delta$-degrees of freedom) of
Ref.~\cite{Piarulli:2014bda} explicitly violates Weinberg's counting
since it has N2LO long distance and N3LO short distance pieces, and
residuals are not gaussian. More recently, the local short distance
components of this potential have been fitted up to 125 MeV LAB
energy~\cite{Piarulli:2016vel} improving the goodness of the fit,
similarly to~\cite{Perez:2013za} (see also table
\ref{tab:Cut-offdependence}).

\section{Chiral interactions and zero energy renormalization}

The low energy threshold parameters allow to probe the structure of
chiral potentials against the NN interaction. The current approach to
chiral interactions is to incorporate the $\chi$TPE tail and include
short range counterterms fitted to pp and np phase-shifts or
scattering data~\cite{Ekstrom:2013kea,Ekstrom:2014dxa}.~\footnote{In
  momentum space counterterms corresponds to coefficients of
  polynomials, see e.g.  \cite{RuizArriola:2016vap}, which can be
  fixed by low energy threshold parameters by implicit
  renormalization.}  However, these approaches are subject to strong
systematic uncertainties since a fit to phase-shifts may be subjected
to off-shell ambiguities and so far low energy chiral potentials
fitted to data have not achieved gaussian
residuals~\cite{Ekstrom:2014dxa} or even have
huge~\cite{Gezerlis:2014zia} or moderate \cite{Piarulli:2014bda}
$\chi^2/\nu$ values. To avoid these shortcomings we use
$\chi$TPE~\cite{Perez:2013oba,Perez:2013cza} with a simpler short
range structure inferred from low energy threshold
parameters~\cite{Perez:2014waa} with their uncertainties inherited
from the 2013-Granada fit~\cite{Perez:2013jpa}. This corresponds to
zero energy renormalization condition of the counterterms.

\begin{table}[h]
\centering
\caption{\label{tab:LEDeltaShellsTPE} Delta-Shell parameters located
  at $r_1=0.9$fm and $r_2=1.8$fm reproducing low energy threshold
  parameters for the indicated waves in the DS-$\chi$TPE potential
  (see main text).}
%\label{tab-1}       % Give a unique label
% For LaTeX tables you can use
\begin{tabular}{cccccccccccc}
\hline
    &$^1S_0$&$^3P_0$& $^1P_1$&$^3P_1$& $^3S_1$&$\epsilon_1$&$^3D_1$&$^3P_2$\\\hline
$\lambda_1$&-0.572(7)& $-$ & 
$-$ & $-$ &-0.368(9)&-0.706(7)& -4.15(1)&$-$ 
 \\
$\lambda_2$& -0.201(3) & -0.033(3) & 
0.103(7) & 0.221(2) & -0.246(4) & 
-0.386(7) & 0.35(1)   &  -0.125(1) 
\\
\hline 
\end{tabular}
% Or use
%\vspace*{5cm}  % with the correct table height
\end{table}

\begin{figure*}[h]
\centering
\includegraphics[width=\linewidth]{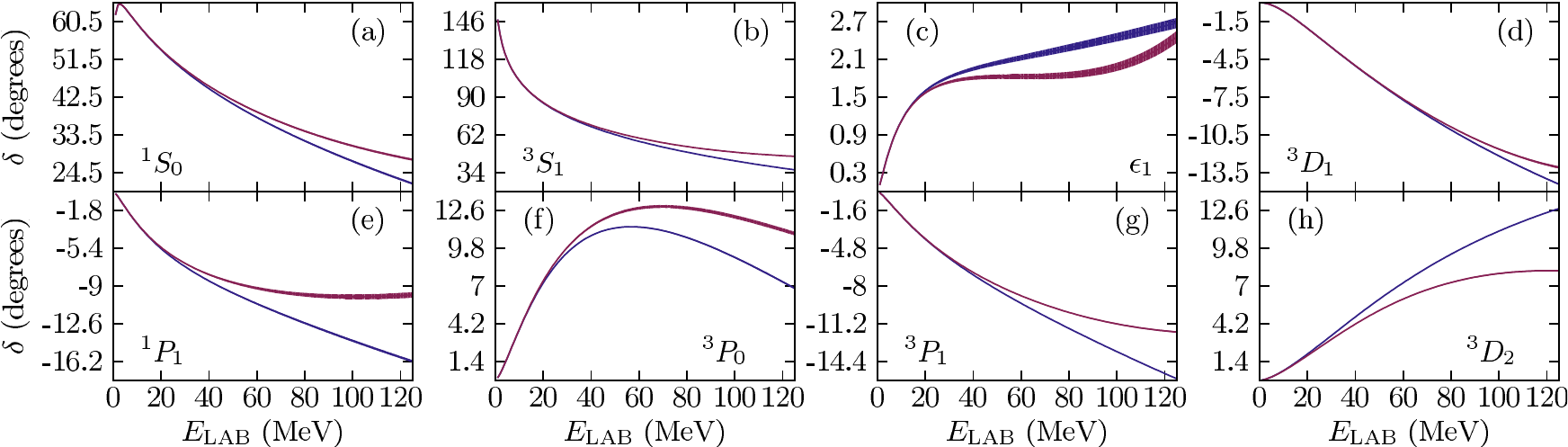}
\caption{$\chi$TPE zero energy renormalized np phase-shifts fixing the
  low energy threshold parameters (see main text)~\cite{Perez:2014waa}
  compared with the phases obtained from the fit to the 2013-Granada
  database~\cite{Perez:2013jpa}.}
\label{fig:2deltaPhases}       % Give a unique label
\end{figure*}

One could naively expect to be able to set any number of short range
counterterms to reproduce the same number of low energy threshold
parameters. Actually, in order to have as the 9 counterterms dictated
by Weinberg to N2LO as in~\cite{Ekstrom:2013kea} we need to fix
$\alpha_0$ and $r_0$ for both $^1S_0$ and $^3S_1$ waves, the mixing
$\alpha_\epsilon$ and $\alpha_1$ for the $^3P_0$, $^3P_1$,
$^3P_2$,$^1P_1$~\cite{Perez:2014waa}. In practice this turned out to
be unfeasible in particular for the $J=1$ coupled channel where one
has matrices $\mathbf{a}$ and $\mathbf{r}_0$. If instead one includes
two counterterms in each partial wave in the $J=1$ coupled channel it
is then possible to reproduce the coupled channel $\mathbf{a}$ and
$\mathbf{r}_0$ matrices.  With this structure we have a
total of 12 short range parameters set to reproduce 12 low energy
threshold parameters from~\cite{Perez:2014waa}, and not the 9 expected
from N2LO~\cite{Ekstrom:2013kea}.  Statistical uncertainties can be
propagated by making fits to each of the $1020$ sets of threshold
parameters that were calculated from the bootstrap generated DS
potentials~\cite{Perez:2014jsa}; this directly takes into account any
statistical correlation between low energy
parameters. Table~\ref{tab:LEDeltaShellsTPE} lists the resulting
12-$\lambda_i$ parameters. In Figure~\ref{fig:2deltaPhases} we show
the phase-shifts corresponding to the DS-$\chi$TPE potential with the
parameters of Table~\ref{tab:LEDeltaShellsTPE} and compare them to the
DS-OPE potential~\cite{Perez:2013mwa, Perez:2013jpa}. We observe a
good agreement between both representations up to a laboratory energy
of $20$ MeV. 
%For waves with two parameters reproducing both the
%scattering length and effective range the agreement goes up to $50$ or
%$60$ MeV, except for the mixing angle. 

\section{Conclusions}

Chiral nuclear forces have been massively implemented in Nuclear
Physics in the last 25 years with the legitimate hope of providing a
unified description of nuclear phenomena more rooted in QCD and less
model dependent than most of the phenomenological approaches. This
huge effort proves that they are not only calculable but also that
they can be used in light nuclei studies, but their indispensability
remains to be established. Their systematic uncertainties may be large
and they might not be necessarily {\it more predictive} than the usual
phenomenological and non-chiral approaches. Within the EFT approach
there is a residual model dependence regarding the finite cut-off
regularization scheme, which seems to dominate the uncertainties.
Therefore, efforts should be placed on reducing this largest source of
error.  We stress that none of these results invalidates $\chi$TPE
above $r_c=1.8 {\rm fm}$, but it does question the status of 
Weinberg's power counting encoding short distance ignorance.

%\bibliography{biblio}
%\bibliographystyle{utphys}

%\end{document}

\end{document}